\documentclass[10 pt,conference]{IEEEtran}
\IEEEoverridecommandlockouts

\usepackage{cite}
\usepackage{amsmath,amssymb,amsfonts}
\usepackage{algorithmic}
\usepackage{graphicx}
\usepackage{textcomp}
\usepackage{subcaption}
\usepackage{epstopdf}
\usepackage{multicol}
\usepackage{multirow}
\usepackage{siunitx}
\usepackage{float}
\usepackage{footnote}
\usepackage{draftwatermark}
\SetWatermarkText{\Huge{Camera Ready IEEE WCNC 2019}}
\usepackage[table,xcdraw]{xcolor}
\DeclareMathOperator*{\argmax}{arg\,max}

\def\BibTeX{{\rm B\kern-.05em{\sc i\kern-.025em b}\kern-.08em
		T\kern-.1667em\lower.7ex\hbox{E}\kern-.125emX}}

\begin{document}
	
	\title{Sampling Free TDOA Localization in Millimeter Wave Networks}
	\author{\IEEEauthorblockN{Shree Prasad M.$^{*}$, Trilochan Panigrahi$^{*}$, Mahbub Hassan$^{\dagger}$, Ming Ding$^{\ddagger}$}
		\IEEEauthorblockA{$^{*}$Dept. of Electronics and Communication Engineering, National Institute of Technology Goa, India \\
			$^{\dagger}$School of Computer Science and Engineering, University of New South Wales, Sydney, Australia\\
			$^{\ddagger}$ Data61, CSIRO, Sydney, Australia\\
			Email: shreeprasadm@gmail.com, tpanigrahi@nitgoa.ac.in, mahbub.hassan@unsw.edu.au, Ming.Ding@data61.csiro.au}
		}
		\maketitle
		\begin{abstract}
			Time difference of arrival (TDOA) is a widely used technique for localizing a radio transmitter from the difference in signal arrival times at multiple receivers. For TDOA to work, the individual receivers must estimate the respective signal arrival times precisely, which requires sampling the signal at least double the rate of its highest frequency content, commonly known as the Nyquist rate. Such sampling is less practical for the millimeter wave band comprising of frequencies in 30-300 GHz range. In this paper, we propose an energy detection architecture for accurately estimating the time of arrival from a single picosecond Gaussian pulse, which enables TDOA localization without sampling at the receiver. We derive the closed form expression of the estimated time of arrival and validate it via simulation. We demonstrate that the proposed sampling-free TDOA can localize millimeter wave transmitters as accurately as the conventional sampling-based TDOA. 
		\end{abstract}
		\maketitle
		
		\section{Introduction}
		Millimeter wave (mmWave), a newly released spectrum band in the frequency range of 30-300 GHz, is fast becoming a popular choice of networking for a range of applications \cite{b1}. The huge bandwidth of mmWave opens up new opportunities for not only high data rate applications but also high precision localization \cite{b2,b3}. With ultra-wide bandwidth, mmWave nodes can emit a very short pulse on the order of picoseconds for precise time resolution at the receiver \cite{b4}. The arrival times recorded at multiple base stations (BSs) then can be used by the widely used algorithm, called time difference of arrival (TDOA), to localize the mmWave transmitter.  
		
		While TDOA has been used successfully for existing frequency bands, it poses a serious implementation challenge for the mmWave band.  Conventional TDOA requires the BSs to sample the incoming signal at least at Nyquist rate, which is twice the rate of the highest frequency content of the signal. Thus, for mmWave, TDOA would require at least 600 GHz sampling, or even higher if more precise localization is required. Such a high sampling rate cannot be achieved with existing analog-to-digital converters (ADCs) \cite{b5}. 
		
		In this paper, we propose an energy-detection-based TDOA that obviates the need for sampling. In the proposed architecture, time of arrival (TOA) is precisely estimated through a bank of time-delayed energy detectors, which estimate the reception of the pulse (energy) at progressively narrower intervals through many iterations enabled by the delay circuits. By dimensioning the delay bank, desired TOA and TDOA localization accuracies can be achieved. 
		
		We obtain closed-form expression of the estimated TOA as a function of the delay bank parameters. The validity of the closed-form expression is achieved via numerical simulations, which confirm that the estimated TOA converges to the true value with increasing iterations (energy detection banks with more elements). By simulating TDOA for many randomly located nodes within a 2mx2m area, we demonstrate that the proposed sampling-free TDOA can localize millimeter wave transmitters as accurately as the conventional sampling-based TDOA.
		
		The rest of the paper is structured as follows. Related work is presented in section \ref{s:rel}, followed by the system model in section \ref{s:model}. Simulation results are presented and discussed in Section \ref{s:sim}. We conclude the paper in Section \ref{s:con}.
\section{Related Works}
\label{s:rel}
Different localization techniques such as direction of arrival (DOA), TOA, and received signal strength have been used in Impulse Radio (IR) communication systems \cite{b6,b7}. The estimation of DOA of source node transmitting higher order Gaussian pulses in the higher frequency band of mmWave (100 - 325 GHz) and in terahertz band (0.1 - 10 THz) is studied in \cite{b8,b9,b10,b10_1}. In \cite{b11}, TOA estimation technique is investigated for different IR mmWave pulse waveforms (with center frequency 28 GHz and 78 GHz). Both of these DOA and TOA estimation techniques require the IR pulse to be sampled beyond the Nyquist rate, which is not possible with the available state of the art data converters. Hence, sub-Nyquist sampling rate TOA estimation schemes based on low complexity energy detectors is investigated in \cite{b12,b13}, but they require a large number of pulses in order to improve the signal-to-noise ratio (SNR) to provide accurate position estimates \cite{b14}. To our knowledge, TDOA estimation from a single pulse without requiring sampling has not been addressed in the literature.

The closest work in the literature is the proposal of a synchronization architecture for pulse-based communication in the terahertz band (0.1-10 THz) \cite{b20}. The authors of \cite{b20} used an electronically tunable voltage controlled delay (VCD) in conjunction with continuous time moving average (CTMA) detectors. CTMAs allow detection of the pulse in a given interval and the VCDs help narrow down the interval successively with multiple pulse transmissions in the preamble. This architecture allows precise TOA estimation without sampling. We build on the core principles of VCD-CTMA-based TOA estimation as proposed in  \cite{b20}, but it is extend for TDOA estimation with a single pulse.

\section{System Model}
\label{s:model}

\begin{figure}
	\vspace{-2mm}
	\centering
	\includegraphics[width=0.9\columnwidth, height = 6cm]{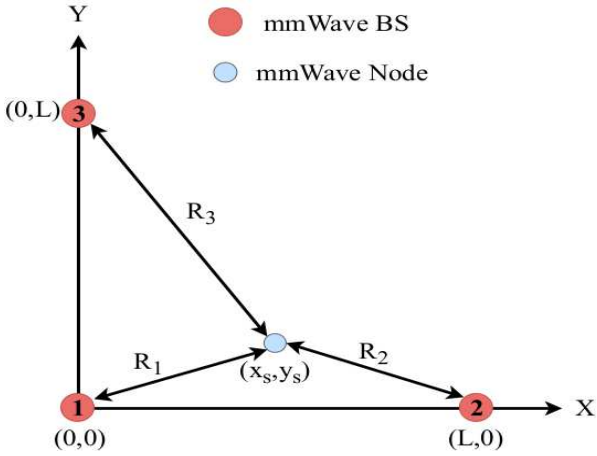}
	\caption{mmWave Network Topology.}
	\label{fig:TOPT}
\end{figure} 
\subsection{Network Topology}
The network topology consisting of three mmWave BSs for TDOA based localization of a mmWave node is shown in Fig. \ref{fig:TOPT}.  BS 1 is located at the origin and the position coordinates of BS 2 and 3 are $\left(0, L \right) $ and $\left(L, 0 \right) $, respectively. The coordinates of mmWave node to be localized is $\left(x_{s},y_{s} \right)$. Both the BSs and the mmWave node employ based graphene-based plasmonic transceivers, which have been recently shown to support efficient wireless communications at very high frequencies above 100 GHz \cite{b20_1}. The mmWave node transmits higher order Gaussian pulses and the TDOA is achieved by estimating the TOA of the pulse at the BSs. 

\subsection{Channel Model}
mmWave channel is still not known very well, especially for the very high frequency range above 100 GHz. Most work in the literature is  focused on lower ranges of the mmWave band (below 60 GHz) using existing channel models.  However, as electromagnetic waves at frequencies over 100 GHz coincide with the resonance frequencies of many molecules in the air, some recent works used radiative transfer theory to model such high frequency channels \cite{b21}, which is used in our system. Channel response and noise of this model are explained below.

%\subsubsection{Channel Response}
The channel response $H\left( f,R_{i}\right)$ accounts for both spreading loss $H_{spread}\left(f,R_{i} \right) $ and molecular absorption loss $H_{abs}\left(f,R_{i} \right) $ and is represented in frequency domain as 
\begin{equation}\label{eq:chresp}
\resizebox{0.5\columnwidth}{!}{$H\left( f,R_{i}\right) = H_{spread}\left( f,R_{i}\right) H_{abs}\left( f,R_{i}\right) $} 
\vspace{-0.5mm}
\end{equation}
\begin{equation}
\resizebox{0.65\columnwidth}{!}{$H_{spread}\left( f,R_{i}\right) =\left(  \frac{c_{o}}{4\pi R_{i} f_{c}}\right) \exp\left( -\frac{j 2\pi R_{i}}{c_{o}}\right)$} 
\vspace{-0.5mm}
\end{equation}
\begin{equation}
\resizebox{0.5\columnwidth}{!}{$H_{abs}\left( f,R_{i}\right) = \exp(-0.5 k\left(f \right)R_{i})$} 
\end{equation}
where $f$ denotes frequency, $R_{i}$ is the path length between $i^{\text{th}}$ mmWave BS and mmWave node, $c_{o}$ is the velocity of light in vacuum, and $k\left( f\right) $ is the medium absorption coefficient. The medium absorption coefficient $k\left( f\right) $ of the mmWave channel at frequency $f$ composed of $J$ type molecules is given as
\begin{equation}
\resizebox{0.3\columnwidth}{!}{$k\left( f\right)  = \sum\limits_{j=1}^{J}a_{j}K_{j}\left( f\right).$}
\end{equation} 
where $a_{j}$ is the mole fraction of molecule type $j$ and $K_{j}$ is the absorption coefficient of individual molecular species.  
%Fig. \ref{fig:MOLE_COEFF} shows the absorption coefficient $k\left(f \right) $ for the mmwave chanel for standard summer air with 1.86\% concentration of water vapor.
%\begin{figure}
%	\vspace{-2mm}
%	\centering
%	\includegraphics[width=0.8\columnwidth, height =3 cm]{mole_coeff.eps}	\caption{Molecular absorption coefficient for standard summer air mmwave channel.}
%	\label{fig:MOLE_COEFF}
%\end{figure} 
The channel response $h\left( t,R_{i}\right)$ is obtained by taking inverse Fourier transform of $H\left( f,R_{i}\right)$. Since, this inverse Fourier transform does not have an analytical closed form expression, the channel response $h\left( t,R_{i}\right)$ is computed numerically.
%\subsubsection{Molecular Absorption Noise}
The ambient noise affecting the propagation of mmWave pulse in mmWave band arises due to the propagating of pulse itself and is defined as molecular absorption noise. The total molecular absorption noise power spectral density (p.s.d.) $S_{N}\left( f,R_{i}\right) $ affecting the transmitted pulse is the sum of background atmospheric noise p.s.d \(S_{N_{B}}\left( f,R_{i}\right)\) and the self induced noise p.s.d. $S_{N_{G}}\left(f,R_{i} \right)$ and is given as\cite{b21}
\begin{equation}
%\vspace{-1mm}
\resizebox{0.27\textwidth}{!}{$S_{N}\left( f,R_{i}\right)  = S_{N_{B}}\left(f,R_{i} \right)+S_{N_{G}}\left(f,R_{i}\right) $} 
\vspace{-1.5mm}
\end{equation}
\begin{equation}\label{eqn:mnm1}
\vspace{-1mm}
\resizebox{0.41\textwidth}{!}{$S_{N_{B}}(f, R_{i}) = \lim\limits_{R_{i} \rightarrow \infty} k_{B} T_{0}\left( 1-\exp\left( -k\left(f \right)R_{i} \right) \right) \left( \frac{c_{0}}{\sqrt{4\pi}f_{c}}\right)^{2}$} 
\vspace{-0.5mm}
\end{equation}
\begin{equation}\label{eqn:mnm2}
\vspace{-0.5mm}
\resizebox{0.40\textwidth}{!}{$S_{N_{G}}\left(f,R_{i} \right) = S_{G}\left( f\right)\left( 1-\exp\left( -k\left(f \right)R_{i}\right) \right) \left( \frac{c_{0}}{4\pi R_{i} f_{c}}\right)^{2} $} 
\vspace{-0.5mm}
\end{equation}
where $k_{B}$ is the Boltzmann constant, $T_{0}$ is the room temperature and $S_{G}\left( f\right) $ represents p.s.d. of transmitted pulse.  

The noise $n\left( t,R_{i}\right) $ effecting the propagation of mmWave pulse is modeled as zero mean additive white Gaussian noise and its power or variance $\sigma^{2}$ is computed as
\begin{equation}
\sigma^{2}\left( R_{i}\right) = \int\limits_{B} S_{N}(f, R_{i}) df
\end{equation}
where $B$ represents the bandwidth of the graphene-based plasmonic transceiver.

Based on this channel model, the pulse received at the mmWave BSs  is represented as \cite{b21}
\begin{equation}
v\left(t,R_{i} \right) = g_{p}\left( t\right)\ast h\left( t,R_{i}\right) + n\left(t,R_{i} \right)  
\end{equation}
\begin{equation}
R_{i} = \tau_{i} c_{o}
\end{equation}
%where $R_{i}$ is the path length or distance between mmWave node and $i^{\text{th}}$ mmWave BS and $\tau_{i}$ is the corresponding true TOA, $g_{p}\left( t\right)$ represents time derivative of a Gaussian pulse with derivative order $p$, $h\left( t,R_{i}\right)$ is the mmWave channel response and $n\left(t,R_{i} \right) $ is the molecular absorption noise created between mmWave node and $i^{\text{th}}$ mmWave BS. The mmWave channel response and molecular absoprtion noise effecting the propagation of mmWave is modeled using radiative transfer theory \cite{b21} and is described in the following subsections.
where $\tau_{i}$ is the true TOA between mmWave node and $i^{\text{th}}$ mmWave BS, $g_{p}\left( t\right)$ represents time derivative of a Gaussian pulse with derivative order $p$, $h\left( t,R_{i}\right)$ is the mmWave channel response and $n\left(t,R_{i} \right) $ is the molecular absorption noise created between mmWave node and $i^{\text{th}}$ mmWave BS.

%Based on the characteristics of the mmwave channel and molecular absorption noise, it is observed that the accuracy of TOA estimate depends upon the center frequency, bandwidth, signal-to-noise (SNR) and the distance L.
\subsection{CTMA Pulse Detector vs. Sampling}
\begin{figure}
	\vspace{-2mm}
	\centering
	\includegraphics[width=\columnwidth, height =5cm]{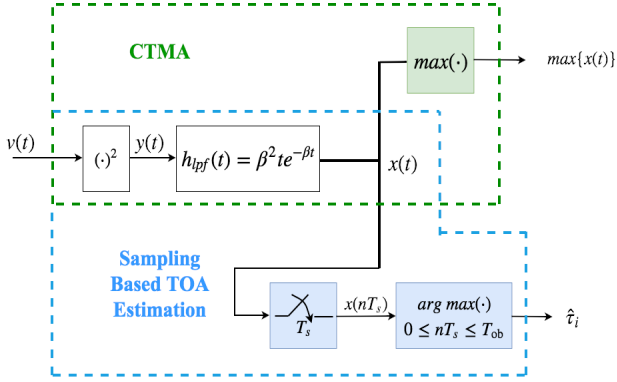}
	\caption{CTMA pulse detector and sampling based TOA estimation.}
	\label{fig:CTMA_BLOCK_V1}
\end{figure}

Fig. \ref{fig:CTMA_BLOCK_V1} illustrates the difference between CTMA pulse detection and sampling. The CTMA pulse detector is modeled as a linear time-invariant system and its input to output relation is given as\cite{b21}:

%
%This section reviews the CTMA pulse detector for IR pulse based communication in 0.1 - 10 THz frequency band . Fig. \ref{fig:CTMA_BLOCK_V1} shows the block diagram of the CTMA pulse detector.  
\begin{equation}\label{eqn:ctma}
x\left(t \right)  = \int\limits_{t-T_{p}}^{t} v^{2}\left( t\right) dt
\end{equation}
where $x\left(t \right)$ is the output of CTMA, $T_{p}$ is the total pulse duration of higher order Gaussian pulse and $v\left(t \right)$ is the input to the CTMA. Equation \eqref{eqn:ctma} can be approximated by a second order low pass filter (LPF) with impulse response $h_{lpf}\left( t\right)  =\beta^{2}te^{-\beta t}$, where $\beta = 1.4615/T_{P}$. Using energy storage elements, such as capacitors, it is possible to hold the maximum value of $x\left(t \right)$ within a certain time interval. A mmWave pulse is detected within the observation time interval $t\in\left[0, T_{\text{ob}}\right] $, if $\text{max}\left\lbrace x\left( t\right)  \right\rbrace $ exceeds a predefined threshold voltage. Thus a CTMA can be used to obtain the $\text{max}\left\lbrace x\left( t\right)  \right\rbrace $, which we will use later in our proposed sampling-free TDOA architecture.
%\subsection{Sampling based TOA Estimation}
%The CTMA pulse detector described in previous section can use a sampler to estimate the TOA of the mmwave pulse at mmwave BS as shown in Fig. \ref{fig:CTMA_BLOCK}. In this method, the output of LPF is sampled above Nyquist rate. Thus, the TOA estimate of the mmwave pulse is the time instant at which the output of LPF is maximum. But this method requires very high sampling rates which is not is possible with the existing state of the art ADCs.
%The LPF employed in CTMA pulse detector described in previous section only provides the maximum value of output of LPF $h_{lpf}\left( t\right)$. can be used to estimate the TOA of the mmwave pulse at mmwave BS as shown in Fig. \ref{fig:CTMA_BLOCK_V1}. In the sampling based TOA estimation method, the output of LPF $h_{lpf}\left( t\right)$ is sampled above Nyquist rate. Hence, the TOA estimate of the mmwave pulse is the time instant at which the output of LPF $h_{lpf}\left( t\right)$ is maximum. But this method requires very high sampling rates which is not is possible with the existing state of the art ADCs. Thus in this paper, a sampling free approach for TOA estimation is proposed by using an array of voltage controlled delays (VCD) and CTMA pulse detector.
%The output of LPF  $h_{lpf}\left( t\right)$ employed in CTMA pulse detector can be sampled to obtain the TOA estimate of mmwave pule as shown in Fig. \ref{fig:CTMA_BLOCK_V1}. 

In the sampling based TOA estimation method, on the other hand, the output of LPF $h_{lpf}\left( t\right)$ is sampled above Nyquist rate and the TOA is estimated as the time instant at which the output of LPF $h_{lpf}\left( t\right)$ is maximum. But this method requires very high sampling rates which is not possible with the existing state of the art ADCs. Thus in this paper, a sampling free approach for TOA estimation is proposed by using an array of VCDs and CTMA pulse detectors.

\subsection{Proposed Sampling free TDOA Estimation}

%of the single mmWave pulse transmitted by mmwave node is obtained by modifying the synchronization scheme proposed in \cite{b20}, which uses multiple pulses in its preamble. Graphene based voltage controlled delay (VCD) is used to delay the pulse and the delay introduced by these VCDs can be easily tuned by means of bias voltage.
 \begin{figure*}[h]
	\centering
	\includegraphics[width=\textwidth, height = 5.5cm]{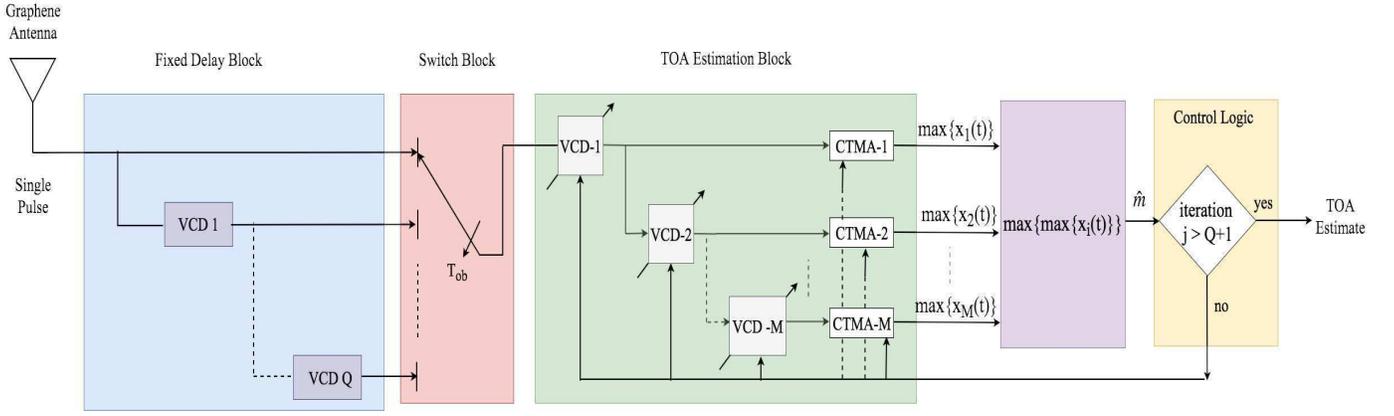}
	\caption{Proposed mmWave BS architecture for TOA Estimation using single mmWave pulse.}
	\label{fig:TOA_EST}
\end{figure*} 
\begin{figure}[h]
	\vspace{-2mm}
	\centering
	\includegraphics[width=\columnwidth, height = 5.5cm]{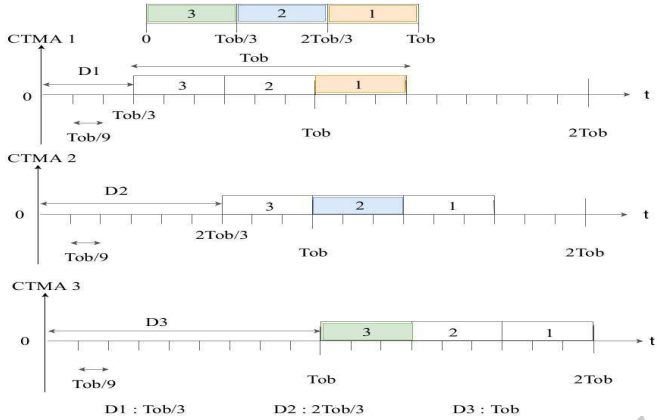}
	\caption{Timing diagram illustrating observation time interval selected by CTMAs for $M = 3$ at first iteration with $T_{\text{win}} = T_{\text{ob}}/3$.}
	\label{fig:ITR1}
\end{figure} 
TDOA requires TOA estimations (TDOA details explained in Section \ref{s:tdoa}). Fig. \ref{fig:TOA_EST} shows the proposed scheme to obtain the TOA estimate at the mmWave base station (BS). In our architecture, there are two delay blocks. The VCDs in the fixed delay block delays the mmWave pulse by $T_{\text{ob}}$ and thus provide a copy of the single mmWave pulse to the next block ever $T_{\text{ob}}$ seconds. In contrast, the delay introduced by VCDs in TOA estimation block is tuned for every $T_{\text{ob}}$. Without loss of generality, the procedure for obtaining the TOA estimate at a mmWave BS $i$ is described as follows. Here, it is assumed that the mmWave pulse arriving at all the three mmWave BSs is contained within $T_{\text{ob}}$.

During the first iteration, TOA of mmWave pulse is estimated within $T_{\text{ob}}$. The delay introduced by VCDs in TOA estimation block is $T_{\text{ob}}/M$ and observation window duration $T_{\text{win}}$ of CTMA is $T_{\text{ob}}/M$. Hence, $m_{i}^{\text{th}}$ CTMA computes its output $y\left(t \right) $, for $t\in\left[ \left( M-m_{i}\right)T_{\text{ob}}/M, \left(M-\left( m_{i}-1\right) \right)T_{\text{ob}}/M\right]$. Thus, together all the CTMAs computes its output in contiguous time intervals over entire $T_{\text{ob}}$. In the proposed scheme, the CTMA that yields the maximum output when compared to other CTMAs is said to have detected the mmWave pulse. In general, the decision rule for selecting the CTMA at $j^{\text{th}}$ iteration is given as 
\begin{equation}
\hat{m}_{i,j} = \argmax_{0\leq m_{i}\leq M}\:\: \left\lbrace\text{max}\left\lbrace  x_{m_{i}}\left( t\right)\right\rbrace \right\rbrace  
\end{equation}

The timing diagram in Fig.\ref{fig:ITR1} illustrates the observation time interval selected by CTMAs for $M = 3$ at the first iteration. 
% by dividing Tob into contiguous time slots of duration Tob/M.
After first iteration, the coarse TOA estimate $\hat{\tau}_{i,1}$ of the pulse at the $i^{\text{th}}$ mmWave BS is given as
\begin{equation}
\hat{\tau}_{i,1} = \frac{\left(M-\hat{m}_{i,1}\right) T_{\text{ob}}}{M}
\end{equation}
where $\left(M-\hat{m}_{i,1}\right) T_{\text{ob}}/M$ is the start time interval of the $\hat{m}_{i,1}$ CTMA pulse detector that yielded the maximum output value when compared to the other CTMAs. It should be noted here that, the start time of mmwave pulse can computed as $\hat{\tau}_{i,1}-T_{p}$. Further, either $\hat{\tau}_{i,1}$ or $\hat{\tau}_{i,1}-T_{p}$ can be used in TDOA based localization.
				 
The coarse TOA estimate obtained from the first iteration can be made finer if more pulses were available. Hence the fixed delay block along with switch block is used to obtain delayed copies of the single mmWave pulse. The switch in the switch block changes its position for every $T_{\text{ob}}$. In the second iteration, the TOA estimate of mmWave pulse is estimated within a narrower time interval $\left[ \left(M-\hat{m}_{i,1}\right)T_{\text{ob}}/M, \left(M-\left(\hat{m}_{i,1}-1 \right) \right)T_{\text{ob}}/M \right] $.
		 
In the second iteration, the single pulse delayed by $T_{\text{ob}}$ is given as input to the TOA block. In this iteration the observation window of the CTMA detector $T_{\text{win}}$ is set to $T_{\text{ob}}/M^{2}$. The VCD in the first line introduces a delay of $\left[ \left(\hat{m}_{i,1}-1 \right)T_{\text{ob}}/M + T_{\text{ob}}/M^{2} \right] $ and the remaining VCDs introduces delay of $T_{\text{ob}}/M^{2}$. Thus the total delay in the $m^{\text{th}}$ VCD is $\left(\hat{m}_{i,1}-1 \right)T_{\text{ob}}/M + mT_{\text{ob}}/M^{2}$. Thus, $m_{i}^{th}$ CTMA computes its output $y\left( t\right) $, for $t\in \left[\hat{\tau}_{i,1}+ \left(  M-m_{i}\right)T_{\text{ob}}/M^{2},\: \hat{\tau}_{i,1}+\left(M-\left( m_{i}-1\right) \right)T_{\text{ob}}/M^{2}\right] $. Let $\hat{m}_{i,2} $ represent the CTMA detector that yielded the maximum output value when compared to the other CTMAs. Thus finer TOA estimate $\hat{\tau}_{i,2}$ of the pulse after the second iteration is given as
		 \begin{equation}
		 \hat{\tau}_{i,2} = \hat{\tau}_{i,1}+\frac{\left(M-\hat{m}_{i,2}\right) T_{\text{ob}}}{M^{2}}
		 \end{equation}
Based on the logic described in the first two iterations, the TOA estimate $\hat{\tau}_{i}$, at the $i^{\text{th}}$ mmWave BS after $Q+1$ iterations is given as
		 \begin{equation}\label{eq:main_eq}
		 \hat{\tau}_{i} = \hat{\tau}_{i,Q+1} = \sum\limits_{q=1}^{Q+1}\frac{\left(M-\hat{m}_{i,q}\right) T_{\text{ob}}}{M^{q}}
		 \end{equation}
		 
\begin{figure*}
	\centering
	\begin{subfigure}{\columnwidth}
		\includegraphics[width=\columnwidth, height = 4cm]{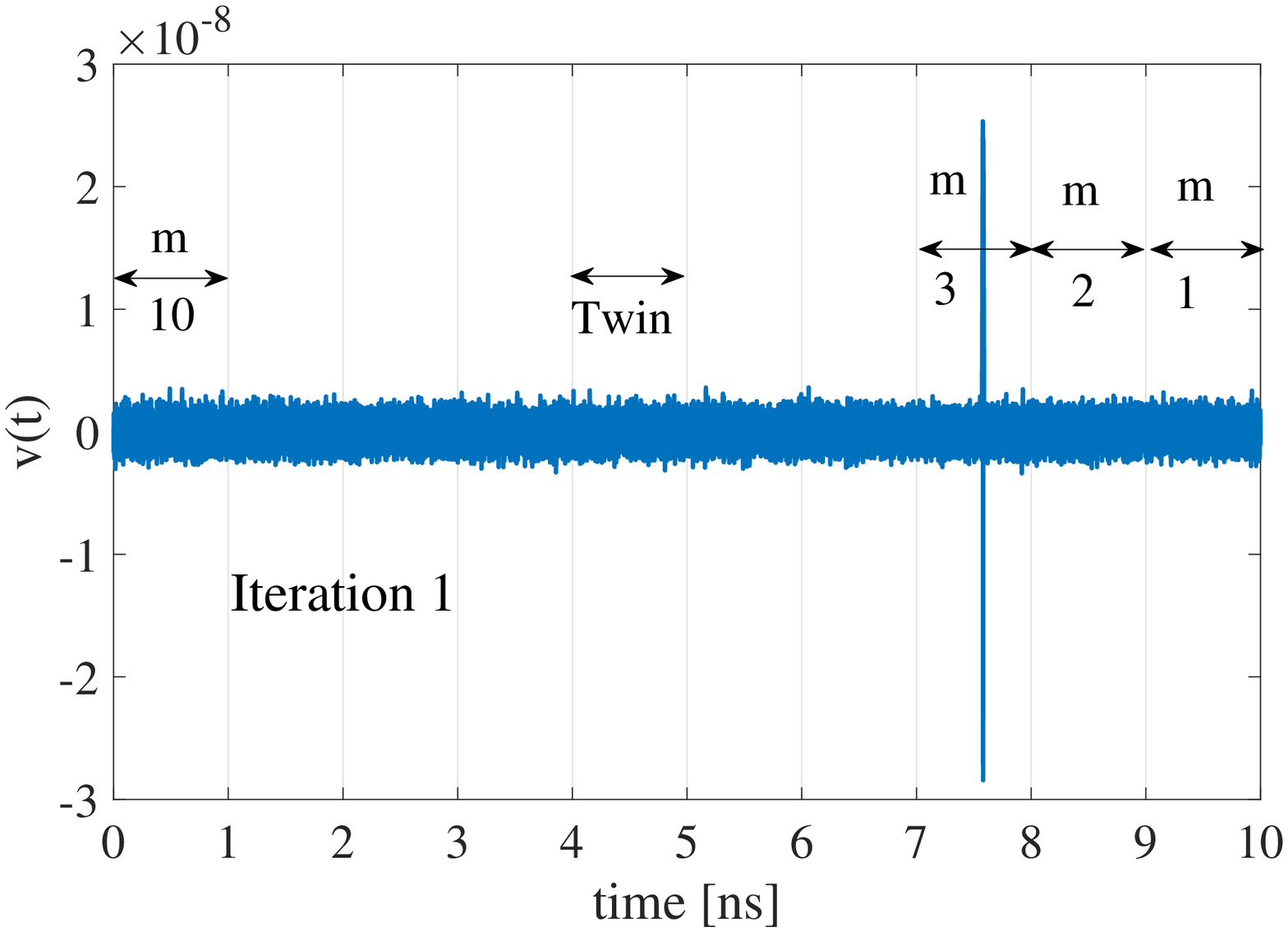}
		\label{fig:CTMA_FIRST}
	\end{subfigure}
	\begin{subfigure}{\columnwidth}
		%\vspace{-6mm}
		\includegraphics[width=\columnwidth, height = 4cm]{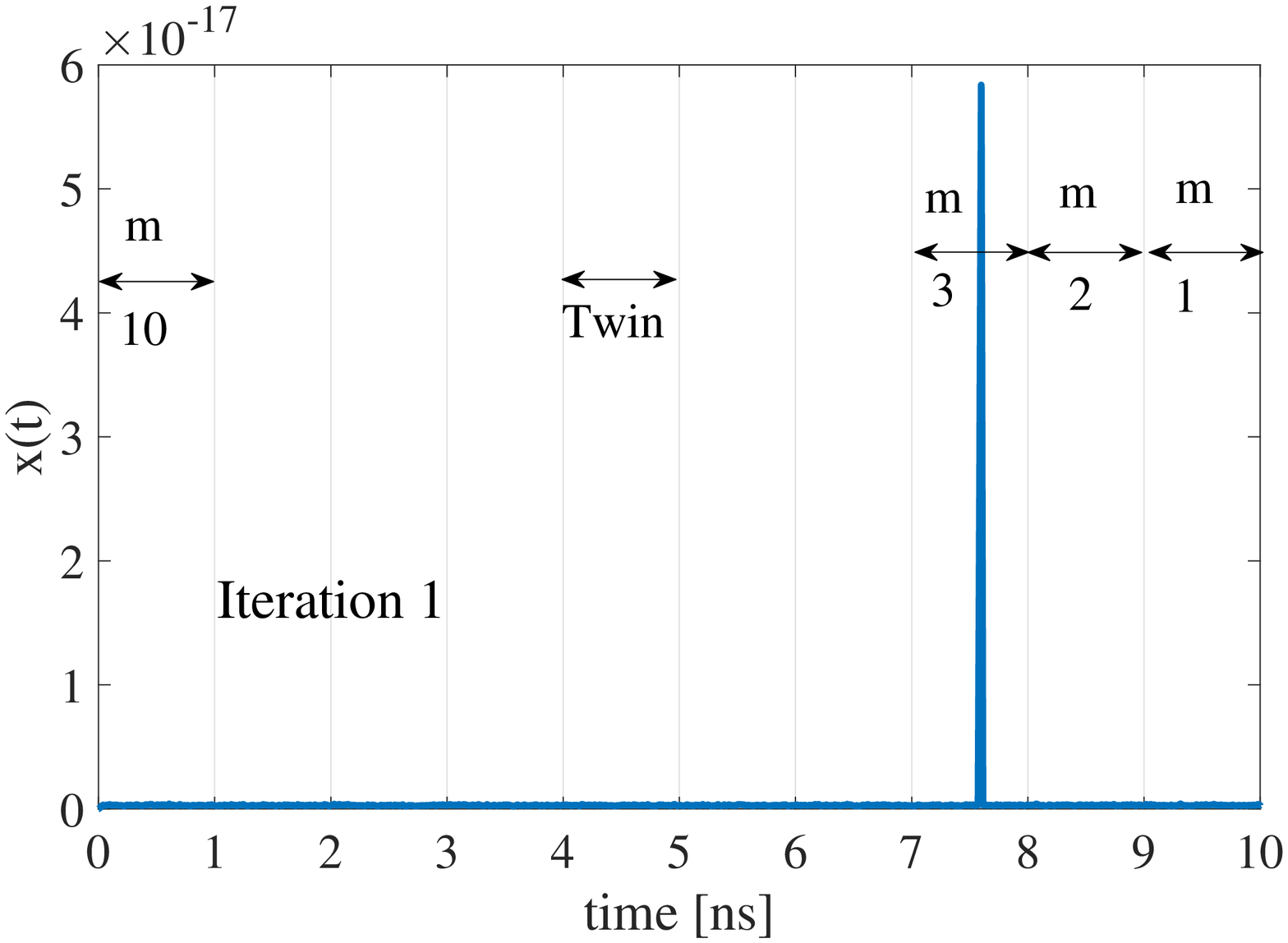}
		\label{fig:PULSE_FIRST}
	\end{subfigure}
\begin{subfigure}{\columnwidth}
	\includegraphics[width=\columnwidth, height = 4cm]{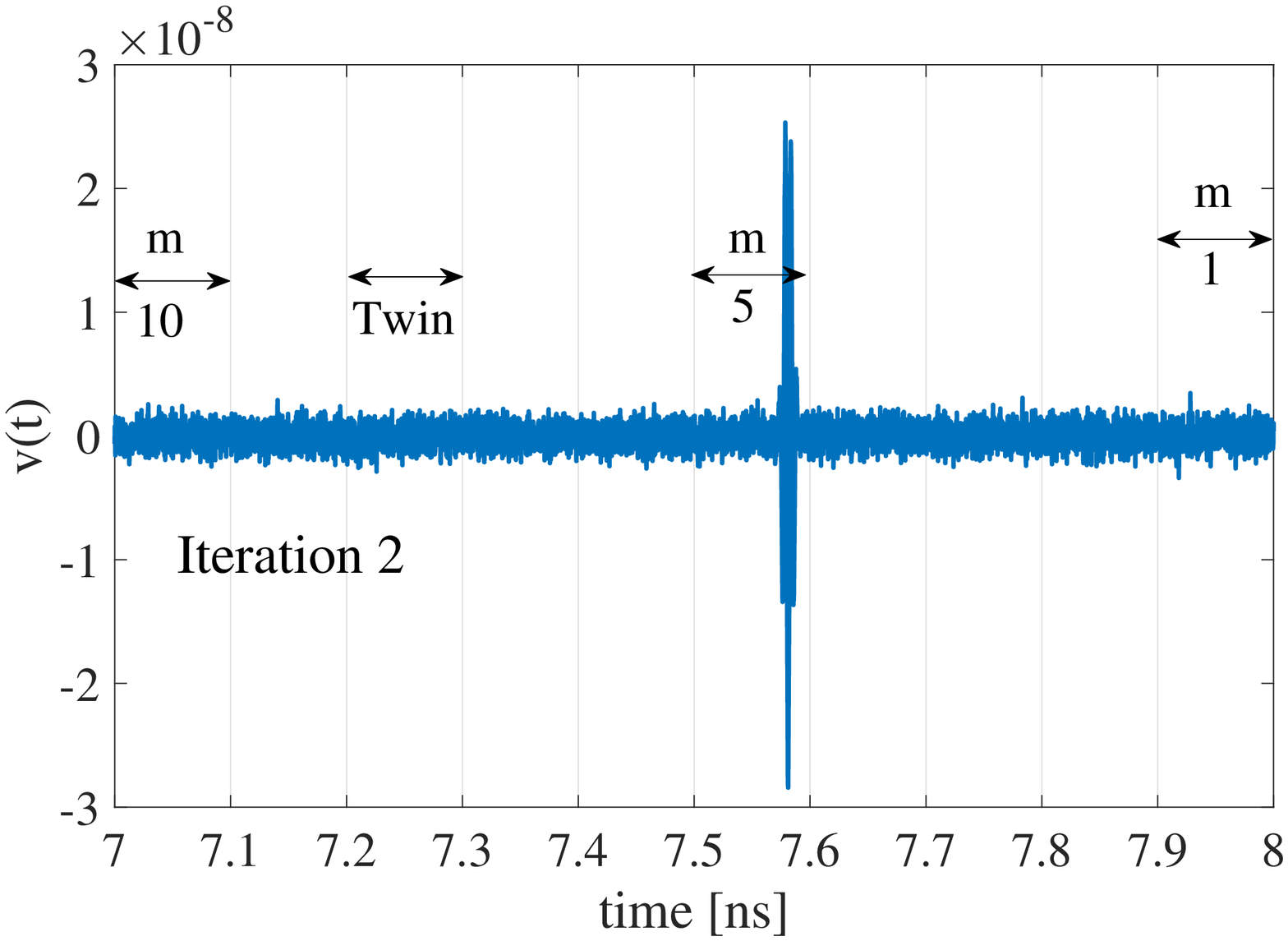}
	\label{fig:CTMA_SECOND}
\end{subfigure}
\begin{subfigure}{\columnwidth}
	%\vspace{-6mm}
	\includegraphics[width=\columnwidth, height = 4cm]{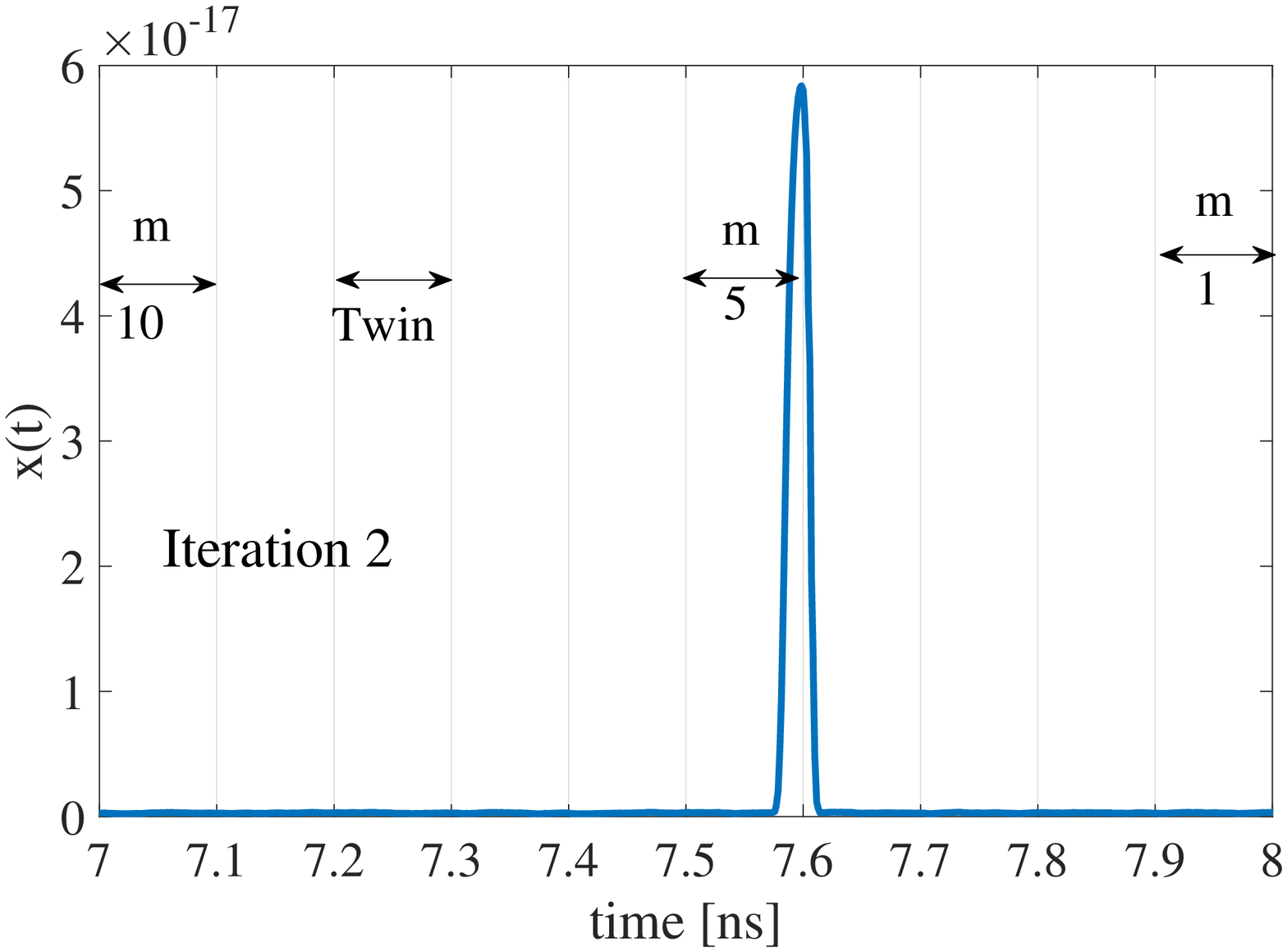}
	\label{fig:PULSE_SECOND}
\end{subfigure}
\begin{subfigure}{\columnwidth}
	\includegraphics[width=\columnwidth, height = 4cm]{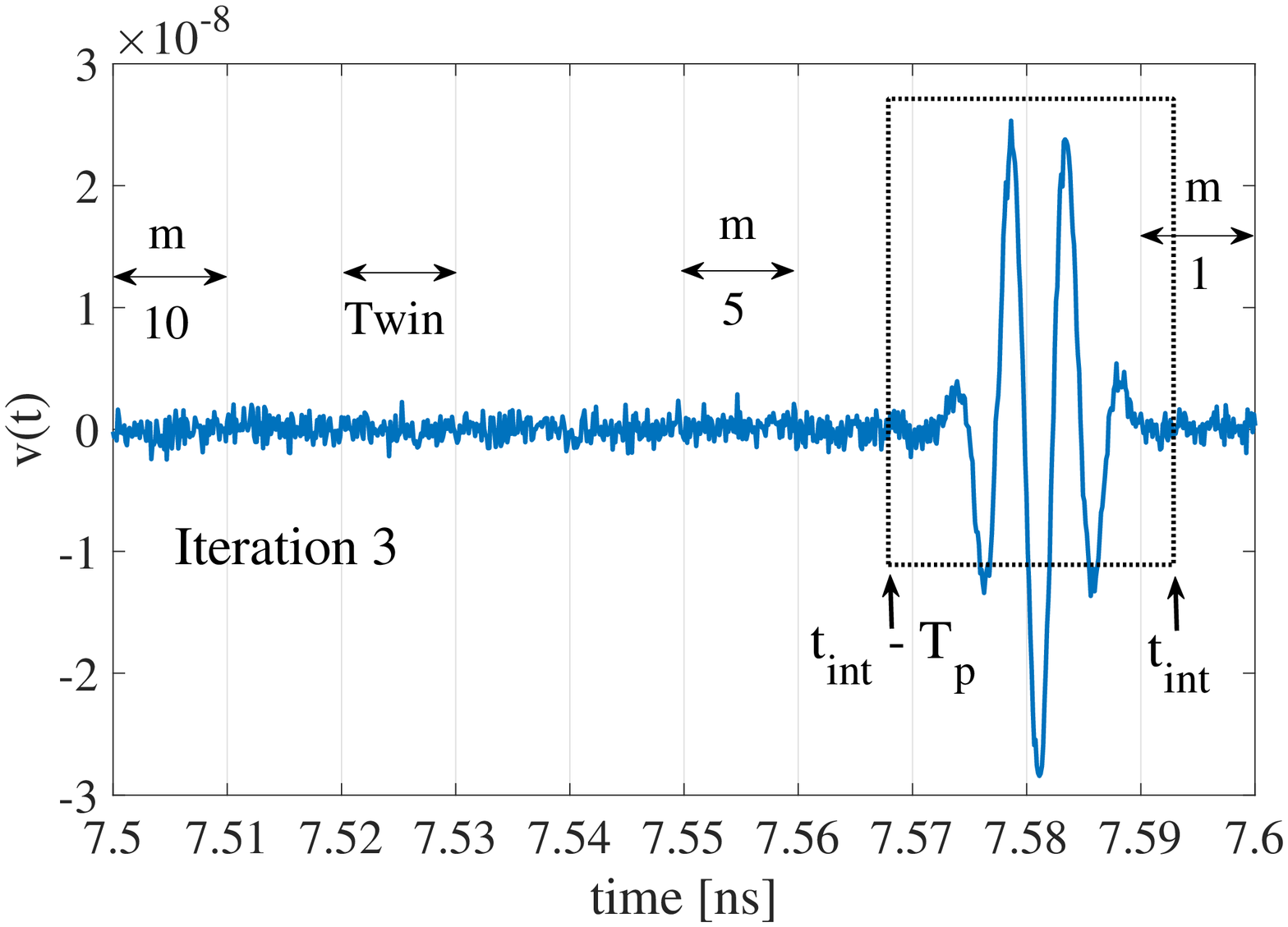}
	\label{fig:CTMA_THIRD}
\end{subfigure}
\begin{subfigure}{\columnwidth}
	%\vspace{-6mm}
	\includegraphics[width=\columnwidth, height = 4cm]{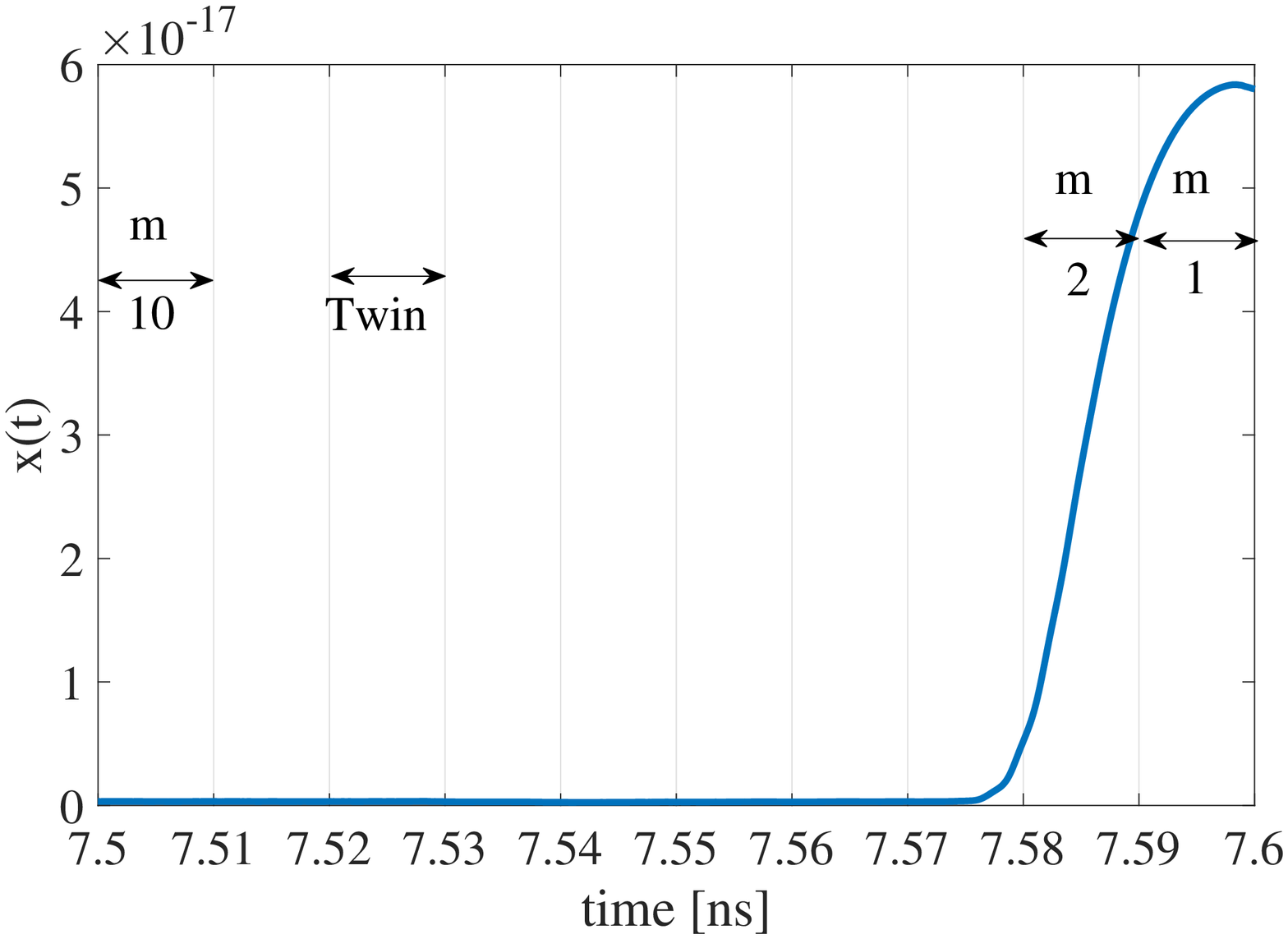}
	\label{fig:PULSE_THIRD}
\end{subfigure}
\caption{Illustration of iterative procedure for estimation of TOA of tenth order Gaussian pulse with center frequency 200 GHz at mmWave BS 1. The number of CTMAs $M = 10$ and fixed VCDs $Q = 2$.}
\label{fig:de_itr}
\vspace{-1mm}
\end{figure*}

From Equation (\ref{eq:main_eq}), we can see that the time resolution of the proposed TDOA architecture is basically governed by $M^{Q+1}$, which is further illustrated in Fig. \ref{fig:de_itr} with a true TOA of 7.5685 ns. In the first iteration, the output of CTMA $m = 3$ will be the maximum and hence the TOA is estimated as $\hat{\tau}_{1,1} = 7\:\text{ns}$. In the second iteration, the TOA is estimated within the narrower interval of $\left[7\:\text{ns},  8\:\text{ns}\right]$. Given that the CTMA $m = 5$ has the maximum output among all CTMAs, TOA estimate at the second iteration is obtained as $\hat{\tau}_{1,1} = 7.5\:\text{ns}$. Thus for the third iteration, the TOA estimation time interval further narrows to $\left[7.5\:\text{ns},  7.6\:\text{ns}\right]$.  In the third iteration, since $T_{\text{win}} < T_{p}$, the mmWave pulse is spread across more than three time slots, but CTMA 1 will yield the maximum output when compared to other CTMAs because, although the time interval for CTMA 1 is $t_{\text{int}}\in\left[7.59\:\text{ns}, 7.6\:\text{ns}\right] $, its convolution integral interval is $\left[t_{\text{int}}-T_{p},t_{\text{int}} \right] $.
%\subsubsection{Latency in TOA Estimation}
%Since VCDs with fixed delay $T_{\text{ob}}$ is used to obtain multiple copies of the single mmWave pulse, their will be latency in estimating its TOA at mmWave BS. The latency $\gamma$ depends on the number of VCDs $Q$ in fixed delay block and the length of observation interval $T_{\tex{b}}$ and is given as
%\begin{equation}
%\gamma =  Q\cdot T_{\text{ob}}
% \end{equation}
% 
% The length of observation duration $T_{\text{ob}}$ depends on longest distance possible between mmwave node and mmwave BS within the area confined by the right isosceles triangle (see Fig. \ref{fig:TOPT})
\subsection{TDOA Localization Algorithm}
\label{s:tdoa}
The TDOA localization algorithm \cite{b22}, uses the TOA estimates of the mmWave pulse at mmWave BS to estimate the position coordinates of the mmWave node. In TDOA localization algorithm all mmWave BS are synchronized by a common clock. The TDOA of mmWave pulse at mmWave BS $i$ with respect to reference mmWave BS 1 located at the origin is given as 
\begin{equation}
\tau_{i1} = \tau_{i} - \tau_{1}
\end{equation} 
where $\tau_{i}$ is the true TOA at $i^\text{th}$ mmWave BS. The distance difference $R_{i1}$ is defined as 
\begin{equation}
R_{i1} = R_{i}-R_{1} = \tau_{i1} c_{o}
\end{equation}
Now, we define the following terms that will be used in obtaining the position estimate of mmWave node.
\begin{equation*}
\boldsymbol{H} = \begin{bmatrix}
 x_{2}&y_{2}\\ x_{3}&y_{3}
\end{bmatrix}, \boldsymbol{c} =
\begin{bmatrix}
-R_{2i}\\-R_{3i}
\end{bmatrix}, \boldsymbol{d} = 0.5\begin{bmatrix}
b_{2}^{2}-R_{21}^{2}\\b_{3}^{2}-R_{31}^{2}
\end{bmatrix}
\end{equation*}
where $\left[x_{i}, y_{i} \right] $ are coordinates of $i^{\text{th}}$ mmWave BS and $b_{i}^{2} = x_{i}^{2}+y_{i}^{2} $. Using these terms, the estimate of position coordinates of the mmWave node is obtained according to the following equation
\begin{equation}
\begin{bmatrix}
\hat{x}_{s}\\\hat{y}_{s}
\end{bmatrix} = \left( \boldsymbol{H}^{T}\boldsymbol{H}\right) ^{-1}\boldsymbol{H}^{T}\left( R_{1}\boldsymbol{c}+\boldsymbol{d}\right) 
\end{equation}
\section{Simulation Results}
\label{s:sim}
The proposed sampling-free TDOA algorithm is implemented in MATLAB 2014a.
 
\subsection{Parameters and Performance metrics}
In the simulation, three mmWave BS with coordinates $\left(0,0 \right) $, $\left(0, L\right) $, and $\left(L,0 \right)$ with $L = 2\:m$ is considered \cite{b11}. The mmWave pulse considered in simulation is a second order Gaussian pulses with its center frequency $f_{c} = 200\:GHz$ and half power bandwidths $f_{l} = 123.38\:GHz$  and  $f_{h} = 288.30\:GHz$. The power of the second order Gaussian pulse is 1 \si{\micro}W. The Molecular absoprtion coefficient $k\left(f \right) $ of the mmWave channel $\left(100 - 300\:\text{GHz} \right) $ for standard summer air with 1.86\% concentration of water vapor is obtained from high-resolution transmission molecular absorption (HITRAN) database \cite{b23}. Further, the performance of the proposed sampling free TDOA estimation algorithm is compared with sampling based method. 
  
We simulate 10 node positions located randomly in the 2mx2m area. For each position, we repeat the simulation 500 times. The TDOA localization accuracy is then defined in terms 
of average Euclidean distance error (AEDE) as follows:
\begin{equation}
\resizebox{0.88\columnwidth}{0.07\columnwidth}{$\text{AEDE} = \frac{1}{N_{pos}}\sum\limits_{i=1}^{N_{pos}}\left\lbrace \frac{1}{N_{run}}\sum\limits_{j=1}^{N_{run}}\sqrt{\left( \left (\hat{x}\left (i, j \right ) -x\left( i\right)  \right )^{2}+\left (\hat{y}\left (i, j \right ) -y\left( i\right)  \right )^{2}\right)}\right\rbrace  $}
\end{equation}
where $N_{pos}$ is the number of random locations, $\left[x\left(i \right),y\left(i \right) \right]$ the true coordinates of the $i^{\text{th}}$ location,  $N_{run}$ the number of times simulation is repeated for each location, and $\left[ \hat{x}\left ( i,j \right ), \hat{y}\left (i, j \right )\right]$ the estimated coordinates of the $i^{\text{th}}$ node in the $j^{\text{th}}$ repetition. For $L = 2\:m$, the observation interval is $T_{\text{ob}} = \left[ 0, 10\:ns\right] $.  Here, $10\:ns > R_{max}/c_{o}$, where $R_{max}$ is the longest possible distance between the mmWave node and mmWave BS within $L\times L$ area. 
\subsection{TOA Estimation}
\begin{figure}
	\vspace{-2mm}
	\centering
	\includegraphics[width=0.9\columnwidth, height = 5cm]{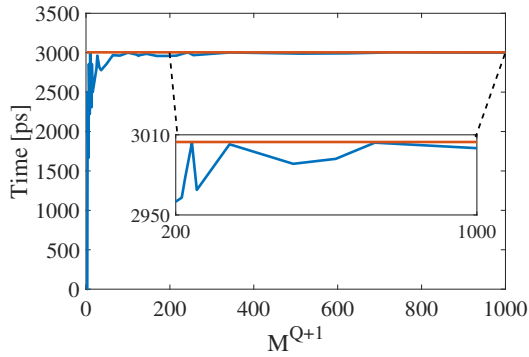}
	\caption{Estimated TOA verus $M^{Q+1}$. Coordinates of the mmWave node transmitting 1 \si{\micro}W power second order Gaussian pulse with $f_{c} = 200\:\text{GHz}$ is $\left[0.5\:m, 0.75\:m \right] $}
	\label{fig:TOA_EST_V}
\end{figure}

The main source of TDOA localization error stems from errors in estimating the TOA of the arriving pulse. As explained by Equation (\ref{eq:main_eq}), TOA accuracy is strictly influenced by $M^{Q+1}$, i.e., the accuracy can be improved by allowing larger $M$ and $Q$ in the architecture. For a node located at $\left[0.5\:m, 0.75\:m \right]$, Figure \ref{fig:TOA_EST_V} plots TOA estimates as a function of $M^{Q+1}$. We make the following observations: (1) with only a single CTMA (m=1) and no iteration (Q=0), the TOA estimation accuracy is very poor, (2) TOA estimate improves very quickly and becomes close to the ground truth with the addition of only a few CTMAs and/or iterations, and (3) the improvement then slows down and it would take many more CTMAs and/or iterations to further reduce the gap between the ground truth and the estimate.
%Fig. \ref{fig:TOA_EST_V} compares the estimated TOA at mmWave BS located at origin using \eqref{eq:main_eq} for different values of $\alpha$ against the true TOA. Here the location of mmWave node is considered as $ \left[0.5\:m, 0.75\:m \right] $. From Fig. \ref{fig:TOA_EST_V} it is observed that the TOA estimated obtained using \eqref{eq:main_eq} approaches the true TOA with increase in value of $\alpha$ and converges to the true TOA for large values $\alpha$.
\vspace{-1mm}
\subsection{Localization Accuracy}
\begin{figure}
 	\vspace{-2mm}
 	\centering
 	\includegraphics[width=0.9\columnwidth, height = 5cm]{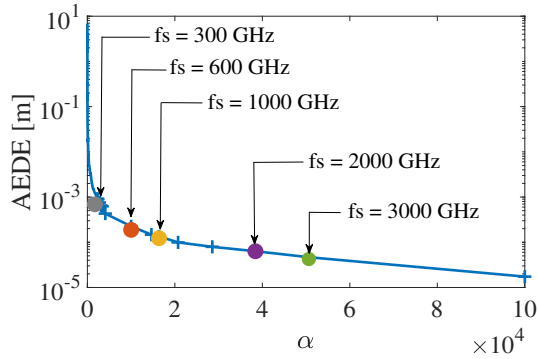}
 	\caption{AEDE [m] versus $M^{Q+1}$. (The AEDE values at different $f_{s}$ and the $M^{Q+1}$ that yields AEDE value closest to the AEDE obtained using different $f_{s}$ is \textbf{i)} $f_{s} = 300\:\text{GHz}, \text{AEDE} = 0.6903\:\text{mm}, \alpha = 1728$, \textbf{ii)} $f_{s} = 600\:\text{GHz}, \text{AEDE} = 0.188\:\text{mm}, \alpha = 4096$, \textbf{iii)} $f_{s} = 1000\:\text{GHz}, \text{AEDE} = 0.123\:\text{mm}, \alpha =  10000$, \textbf{iv)} $f_{s} = 2000\:\text{GHz}, \text{AEDE} = 0.0632\:\text{mm}, \alpha = 16384$, and \textbf{v)}$f_{s} = 3000\:\text{GHz}, \text{AEDE} = 0.0326\:\text{mm}, \alpha = 20736$)}
 	\label{fig:WCNC_ALPHA}
 \end{figure}

Fig. \ref{fig:WCNC_ALPHA} plots AEDE for 10 random locations as a function of $M^{Q+1}$. As expected, the localization error of the proposed TDOA decreases with increasing $M^{Q+1}$ in a similar way the TOA improves in Figure \ref{fig:TOA_EST_V}. To compare the performance of the proposed sampling-free TDOA against the conventional sampling-based method, we also plot the localization errors achieved by the sampling-based method for different sampling rates starting from 300 GHz. We can see that for Nyquist rate equal to 600 GHz, the proposed TDOA can match the performance of sampling-based TDOA with $M^{Q+1}=4096$, which could be achieved for example with only 2 CTMAs subjected to 11 iterations (Q=10). The proposed TDOA can match the sampling-based TDOA for higher than Nyquist-rate samplings by simply increasing the number of CTMAs and/or iterations.   
\section{Conclusion}
\label{s:con}
We have proposed an energy-detection-based TDOA architecture that obviates sampling requirement at the receiver. The proposed TDOA method, therefore, is more suitable for high frequency bands, such as mmWave band. The TOA accuracy of the proposed architecture can be estimated using closed-form expression. Using numerical simulations, we have demonstrated that the proposed sampling-free TDOA can localize millimeter wave transmitters as accurately as the conventional sampling-based TDOA. 

	\end{document}